\documentclass[11pt]{article}
\usepackage{epsfig}
\def\text{}
\renewcommand{\baselinestretch}{1.75} 

\newcommand{\al}{\alpha}

\newcommand{\la}{\lambda}

\def\tshalf{ {\textstyle {1\over 2}} }
\def\ha{ {\textstyle {1\over 2}} }

\def\xtw{}
\def\xf{}
\def\nha{\ha}
\def\xxt{}

\def\NSone{ {\bigcirc \!\!\!\!\! 1}\,\,}
\def\NStwo{ {\bigcirc \!\!\!\!\! 2}\,\,}
\def\NSthree{ {\bigcirc \!\!\!\!\! 3}\,\,}
\def\NSfour{ {\bigcirc \!\!\!\!\! 4}\,\,}

\newcommand{\be}{\begin{eqnarray}}
\newcommand{\ee}{\end{eqnarray}}

\newcommand{\pr}{\partial}

\newcommand{\np}{\newpage}
\newcommand{\hs}{\hspace}
\newcommand{\vs}{\vspace}
\newcommand{\nl}{\newline}
\newcommand{\nn}{\nonumber}

\newcommand{\RR}{{\rm I\kern-1.6pt {\rm R}}}

\newcommand{\ZZ}{{\rm Z}\kern-3.8pt {\rm Z} \kern2pt}

\newdimen\tableauside\tableauside=1.1ex
\newdimen\tableaurule\tableaurule=0.4pt
\newdimen\tableaustep
\def\phantomhrule#1{\hbox{\vbox to0pt{\hrule height\tableaurule
width#1\vss}}}
\def\phantomvrule#1{\vbox{\hbox to0pt{\vrule width\tableaurule
height#1\hss}}}
\def\sqr{\vbox{%
  \phantomhrule\tableaustep
 
\hbox{\phantomvrule\tableaustep\kern\tableaustep\phantomvrule\tableaustep}%
  \hbox{\vbox{\phantomhrule\tableauside}\kern-\tableaurule}}}
\def\squares#1{\hbox{\count0=#1\noindent\loop\sqr
  \advance\count0 by-1 \ifnum\count0>0\repeat}}
\def\tableau#1{\vcenter{\offinterlineskip
  \tableaustep=\tableauside\advance\tableaustep by-\tableaurule
  \kern\normallineskip\hbox
    {\kern\normallineskip\vbox
      {\gettableau#1 0 }%
     \kern\normallineskip\kern\tableaurule}%
  \kern\normallineskip\kern\tableaurule}}
\def\gettableau#1 {\ifnum#1=0\let\next=\null\else
  \squares{#1}\let\next=\gettableau\fi\next}
\newcommand{\fund}{\tableau{1}}
\newcommand{\Ysymm}{\tableau{2}}
\newcommand{\Yasymm}{\tableau{1 1}}
\begin{document}

\thispagestyle{empty}

\vs*{-25mm}
\begin{flushright}
BRX-TH-473\\[-.15in]
BOW-PH-117\\[-.15in]
HUTP-00/A018\\[-.15in]
hep-th/0006141\\
\vs{8mm}
\end{flushright}
\setcounter{footnote}{0}

\begin{center}
{\Large{\bf  M-Theory tested by  ${\cal N}=2$ 
Seiberg-Witten Theory}}
\renewcommand{\baselinestretch}{1}
\small
\normalsize
\vspace{.1in}
\footnote{Based on a talk 
by H.J. Schnitzer at the 
{\it Workshop on Strings, Duality and Geometry,} \\
\phantom{aaa} 
 University of 
Montreal, March 2000.
}\\

\vspace{.3in}

Isabel P. Ennes\footnote{
Research supported 
by the DOE under grant DE--FG02--92ER40706.}$^{,a}$, 
Carlos Lozano\footnotemark[2]$^{,a}$, Stephen G. Naculich\footnote{
Research supported in part by the National 
Science Foundation under grant 
no.~PHY94-07194 through the \\
\phantom{aaa}  ITP Scholars Program.}$^{,b}$, 
Henric Rhedin$^{c}$, \\
Howard J. Schnitzer\footnote{Permanent address.}
${}^{\!\!\!,\!\!\!}$
\footnote{Research supported in part
by the DOE under grant DE--FG02--92ER40706.\\
{\tt \phantom{aaa} naculich@bowdoin.edu; Henric.Rhedin@celsius.se;
ennes,lozano,schnitzer@brandeis.edu}\\}$^{,a,d}$\\

\vspace{.2in}

${}^{a,4}$Martin Fisher School of Physics\\
Brandeis University, Waltham, MA 02454

\vspace{.2in}

${}^{b}$Department of Physics\\
Bowdoin College, Brunswick, ME 04011

\vspace{.2in}

${}^{c}$Celsius Consultants,
Chalmers Teknikpark\\
S-412 88 G\"oteborg, Sweden

\vspace{.2in}

${}^{d}$Lyman Laboratory of Physics \\
Harvard University, Cambridge, MA 02138

\vspace{.3in}

{\bf{Abstract}} 
\end{center}
\renewcommand{\baselinestretch}{1.75}
\small
\normalsize
\begin{quotation}
\baselineskip14pt
\noindent  

Methods are reviewed for computing the instanton expansion 
of the prepotential for ${\cal N}=2$ Seiberg-Witten (SW) theory
with {\it non}-hyperelliptic curves. These results, when compared
with the instanton expansion obtained from the microscopic Lagrangian,
provide detailed tests of \hbox{M-theory}. 

Group theoretic regularities of ${\cal F}_{\rm 1-inst}$ allow 
one to ``reverse engineer" a SW curve for ${\rm SU}(N)$ 
with two antisymmetric representations
and $N_f\leq 3$ fundamental hypermultiplet representations, 
a result not yet available by other methods. Consistency with 
\hbox{M-theory} requires a curve 
of infinite order. 

\end{quotation}

\np 

\setcounter{page}{1}
\noindent{\bf 1. ~Objectives}
\renewcommand{\theequation}{1.\arabic{equation}}
\setcounter{equation}{0}

\baselineskip20pt

We present a method for 
obtaining precise tests of \hbox{M-theory} using 
\hbox{${\cal N}=2$} Seiberg-Witten (SW)
 supersymmetric (susy) gauge theory \cite{SeibergWitten}. Although 
one believes in \hbox{M-theory}, 
it must nevertheless be subjected to detailed
verification for the same reasons that one 
subjects quantum electrodynamics to  precision tests.  
 In our context, 
\hbox{M-theory} provides SW curves for low-energy 
effective \hbox{${\cal N}=2$} 
susy gauge theories, which in principle allows one to 
compute the instanton expansion of the 
prepotential of the theory. The results
of this calculation must be compared with calculations of 
the instanton contributions to the prepotential 
from the microscopic Lagrangian. It is this comparison which provides 
the tests we are concerned with. It should be noted 
that what we are considering is the ability of \hbox{M-theory} 
to make detailed non-perturbative predictions for field 
theory, as we consider the limit in which gravity has decoupled.  

M-theory provides SW curves for effective \hbox{${\cal N}=2$} 
susy gauge theories with
hypermultiplets in both the fundamental representation \cite{Mtheory,
Witten} and in higher representations \cite{LandsteinerLopezLowe}. 
Since the (hyperelliptic) curves from the former were initially obtained
from purely field-theoretic considerations, 
we regard 
these as {\it post}-dictions of \hbox{M-theory}, 
though the agreement is gratifying.  In order to obtain 
genuine tests of \hbox{M-theory} we need to consider 
situations for which it is not known how to obtain SW curves 
from field-theoretic arguments alone; for example,  \hbox{${\cal N}=2$}
${\rm SU}(N)$  gauge theory with a hypermultiplet in the symmetric or 
antisymmetric representation 
\cite{LandsteinerLopezLowe}. In such cases \hbox{M-theory} gives the 
only known predictions 
of the relevant SW curves, which 
happen to be {\it non}-hyperelliptic curves. If one can extract 
the instanton expansion for these 
examples, and compare these to results from a microscopic 
calculation, one will have genuine 
tests of \hbox{M-theory}. The problem
we faced is that there were no known methods to obtain the 
instanton expansion. Our solution to this
issue will be our main theme.

\noindent{\bf 2. ~Seiberg-Witten Theory}
\renewcommand{\theequation}{2.\arabic{equation}}
\setcounter{equation}{0}

Consider  \hbox{${\cal N}=2$} susy 
Yang-Mills theory in $d=4$ dimensions, with gauge 
group ${\cal G}$, together with hypermultiplets 
in some representation $R$. This theory can be described 
by a microscopic Lagrangian 
\be
{\cal L}_{\rm micro}=\frac{1}{4 g^2}\, F_{\mu \nu} ^a\, F^{\mu \nu a}\,+
\, {\theta \over 32 \pi^2}\,F_{\mu \nu} ^a\,\tilde F^{\mu \nu a}\,
+\, D_\mu \phi^{+} D^\mu \phi\,+\, {\rm tr} [\phi,\phi^{+}]^2\,\nn \\
+\, \,\,{\rm fermion}\,+\,{\rm hypermultiplet}\,\, {\rm terms}, 
\label{one}
\ee 
with $\mu, \nu =1$ to $4$ and $a=1$ to dim ${\cal G}$.
The field strength $F_{\mu \nu}$ and the scalar field $\phi$ belong 
to the adjoint representation. 
The vacuum is described by the condition 
\be
[ \phi,\phi^{+}]\,=\,0. 
\label{two}
\ee
Rotate  $\phi^a$  to the Cartan 
subalgebra, in which case 
\be
{\rm diag}\,(\phi) \,=\, (a_1, a_2, \ldots)\,,  
\,\,\,\,\,\,\,\,\,{\rm with} \,\,\,\,\sum_i a_i=0.
\label{three}
\ee
If all the $a_i$ are distinct, 
this generically breaks ${\cal G}$ to 
${\rm U(1)}^{\rm rank \,\,{\cal G}}$.
If only $\phi$ acquires a vacuum expectation value (vev), we define
this  as the Coulomb branch, which is our focus.

Seiberg and Witten 
\cite{SeibergWitten}
formulated the exact solution 
of the low-energy description of \hbox{${\cal N}=2$} susy gauge 
theories in terms of an effective (Wilsonian) action 
accurate to two derivatives of the fields, {\it i.e.} 
 \be
{\cal L}_{\rm eff}\,=\,\frac{1}{4\pi} {\rm Im}\left(\int {\rm d}^4\theta
\frac{\pr {\cal F}(A)}{\pr A_i}\bar{A_i}+
\frac{1}{2}\int {\rm d}^2\theta\frac{\pr^2 {\cal F}(A)}{\pr A_i\,\pr A_j}
W^{\al}_i\,W_{\al,j}\right)\,+\,{\rm higher~derivatives}, \nn \\
\label{four}
\ee 
where $A^i$ are ${\cal N}=1$ chiral 
superfields ($i=1 \,\,\,{\rm to}\,\, \,{\rm rank}\,{\cal G}$), 
${\cal F}(A)$ is the holomorphic prepotential, and $W^i$ is the
gauge field strength. In components the effective action is 
\be
{\cal L}_{\rm eff}\,=\,{1\over 4}  
{\rm Im} (\tau _{ij})\, F_{\mu \nu}^i F^{\mu \nu j}\,+\,
{1\over 4} {\rm Re} (\tau _{ij})\, 
F_{\mu \nu}^i \tilde F^{\mu \nu \,j} \nn \\ 
+ \pr_{\mu} (a^+)^j  \pr^{\mu} (a_D)_j\,+\,{\rm fermions}.
\label{five}  
\ee
We define the
order parameters $a_i$ as in (\ref{three}), 
$(a_D)_j =  {\pr {\cal F}(a)\over \pr a_j}$ are the dual
order parameters, and 
\be
\tau_{ij}= {\pr^2{\cal F}(a)\over \pr a_i \pr a_j},
\label{matrix}
\ee
is the coupling matrix. Note that 
${\rm Im} (\tau_{ij}) \geq 0$ for positive kinetic energies. 

The holomorphic prepotential can be expressed in 
terms of a perturbative part and infinite series 
of instanton contributions as 
\be
{\cal F}(A)\,=\, {\cal F}_{\rm classical}(A)\,
+\,{\cal F}_{\rm 1-loop}(A)\,+\,\sum _{d=1}^{\infty} \Lambda^{(2N-I(R))d}
{\cal F}_{\rm d-inst}(A),
\label{six}
\ee
where we have specialized the 
instanton terms to ${\rm SU}(N)$ as an illustration. Due to a
non-renormalization theorem, the perturbative expansion for (\ref{six})
terminates at \hbox{1-loop},  though there is an infinite series of
non-perturbative  instanton contributions.  In (\ref{six}), $\Lambda$ is
the  quantum scale (Wilson cutoff) and $I(R)$ is the Dynkin
index of matter hypermultiplet(s) of representation $R$. 

The Seiberg-Witten data which (in principle) 
allow one to reconstruct the prepotential
are:

\noindent
1) A suitable Riemann surface or algebraic curve, dependent on moduli 
$u_i$, or equivalently on 
\vskip-.1in
the order parameters $a_i$.
\vskip-.1in
\noindent
2) A preferred meromorphic 1-form $\lambda \equiv {\rm SW}$ differential.
\vskip-.1in
\noindent
3) A canonical basis of homology cycles on the surface $(A_k, B_k)$.
\vskip-.1in
\noindent
4) Computation of period integrals
\be
2\pi i a_k=\oint_{A_k}\la, \hs{15mm} 2\pi ia_{D,k}=\oint_{B_k}\la, 
\label{eight}
\ee
where recall 
$a_{D,k}\,=\,{{\pr {\cal F}{(a)}}\over {\pr a_k}}$ 
is the dual order parameter. The program 
is:

\noindent 
i) find the Riemann surface or 
algebraic curve appropriate to the given matter content,
\vskip-.1in
\noindent 
ii) compute the period integrals, 
\vskip-.1in
\noindent 
iii) integrate these to find ${\cal F}(a)$, and 
\vskip-.1in
\noindent 
iv) test against results from ${\cal L}_{\rm micro}$ when possible.  

What classes of SW curves are encountered for simple, classical
groups $({\rm SU}, {\rm SO}, {\rm Sp})$ with matter hypermultiplets 
consistent with asymptotic freedom? 

\noindent
a) hyperelliptic curves
\vskip-.2in
\be
y^2\,+\, 2A(x) y\,+\,B(x)\,=\,0,
\label{BBuno}
\ee
\vskip-.1in
for pure gauge theory + matter hypermultiplets in
the fundamental representation.

\noindent
b) cubic non-hyperelliptic curves
\vskip-.1in
\be
y^3\,+\, 2A(x)y^2\,+\,B(x)y\,+\,\epsilon(x)\,=\,0,
\label{BBdos}
\ee
\vskip-.1in
which occurs for

i) ${\rm SU}(N)$ + 1 antisymmetric +($N_f< N+2)$ fundamental
hypermultiplets. 
\vskip-.1in  
ii) ${\rm SU}(N)$ + 1 symmetric +($N_f< N-2)$ fundamental
hypermultiplets.

\noindent
c) curves of infinite order from 

i) elliptic models, or 
\vskip-.1in
ii) decompactifications of elliptic models.

{\it Example}: ${\rm SU}(N)\,+\,$ 2 antisymmetric and $N_f\leq 4$
fundamental hypermultiplets.  

 The main task in extracting instanton 
predictions from curves such as (\ref{BBdos}) 
 is the computation of the period 
 integrals (\ref{eight}), and the integration of  
 ${\pr {\cal F}(a)/ \pr a_k}$ to obtain ${\cal F}(a)$. 
There are two principal
 (complementary) methods to evaluate the period integrals 
 for {\it hyperelliptic} curves. These are Picard-Fuchs differential
equations for the period integrals
 \cite{PicardFuchs}, and direct evaluation of the period integrals by
asymptotic expansion
 \cite{DHokerKricheverPhong1, DHokerKricheverPhong2, DHokerPhong}.
 
 The problem we face is how to evaluate period integrals 
 \be
\oint \lambda\,=\,\oint {x dy\over y},
\label{eleven}
\ee
for non-hyperelliptic curves 
such as (\ref{BBdos}).  
For the cubic curve, the exact solution is too 
complicated to be useable, while for curves of higher 
order, even exact solutions are not possible. 
Numerical solutions are of no interest, as we 
want to study the analytic behavior of ${\cal F}(a)$ 
on the order parameters.

\noindent{\bf 3. ~M-theory and the Riemann Surface}
\renewcommand{\theequation}{3.\arabic{equation}}
\setcounter{equation}{0}

The seminal work on this subject is by Witten \cite{Witten}, 
who considers IIA string theory 
lifted to \hbox{M-theory}. 
It is
convenient to use the language of IIA theory in 
describing the brane structure. Consider ${\rm SU}(N)$ gauge theory with
either  an antisymmetric or symmetric matter 
hypermultiplet \cite{LandsteinerLopezLowe}. The \hbox{M-theory} picture 
is

\begin{picture}(430,200)(10,10)

\put(100,50){\line(0,1){150}}
\put(220,50){\line(0,1){150}}
\put(340,50){\line(0,1){150}}

\put(370,50){\vector(1,0){30}}
\put(404,48){$x_6$}

\put(170,45){\vector(-1,0){30}}
\put(124,32){${\rm Mirror\ \ image}$}

\put(70,120){\vector(0,1){30}}
\put(69,153){$v$}

\put(216,124){$\otimes$}
\put(225,124){O$6$}

\put(100,80){\line(1,0){9}}
\put(119,80){\line(1,0){9}}
\put(138,80){\line(1,0){9}}
\put(157,80){\line(1,0){9}}
\put(176,80){\line(1,0){9}}
\put(195,80){\line(1,0){9}}
\put(211,80){\line(1,0){9}}

\put(100,100){\line(1,0){9}}
\put(119,100){\line(1,0){9}}
\put(138,100){\line(1,0){9}}
\put(157,100){\line(1,0){9}}
\put(176,100){\line(1,0){9}}
\put(195,100){\line(1,0){9}}
\put(211,100){\line(1,0){9}}

\put(100,161){\line(1,0){9}}
\put(119,161){\line(1,0){9}}
\put(138,161){\line(1,0){9}}
\put(157,161){\line(1,0){9}}
\put(176,161){\line(1,0){9}}
\put(195,161){\line(1,0){9}}
\put(211,161){\line(1,0){9}}

\put(155,120){$\cdot$}
\put(155,125){$\cdot$}
\put(155,130){$\cdot$}
\put(155,115){$\cdot$}

\put(220,90){\line(1,0){9}}
\put(239,90){\line(1,0){9}}
\put(258,90){\line(1,0){9}}
\put(277,90){\line(1,0){9}}
\put(296,90){\line(1,0){9}}
\put(315,90){\line(1,0){9}}
\put(331,90){\line(1,0){9}}

\put(220,152){\line(1,0){9}}
\put(239,152){\line(1,0){9}}
\put(258,152){\line(1,0){9}}
\put(277,152){\line(1,0){9}}
\put(296,152){\line(1,0){9}}
\put(315,152){\line(1,0){9}}
\put(331,152){\line(1,0){9}}

\put(220,170){\line(1,0){9}}
\put(239,170){\line(1,0){9}}
\put(258,170){\line(1,0){9}}
\put(277,170){\line(1,0){9}}
\put(296,170){\line(1,0){9}}
\put(315,170){\line(1,0){9}}
\put(331,170){\line(1,0){9}}

\put(275,120){$\cdot$}
\put(275,125){$\cdot$}
\put(275,130){$\cdot$}
\put(275,135){$\cdot$}
\put(220,10){\makebox(0,0)[b]{\footnotesize\bf {Figure 1}}}
\end{picture}

There are $3$ parallel \hbox{NS 5-branes} with 
$N$ \hbox{D4-branes} suspended 
between each, and an \hbox{O6-plane} 
on the central \hbox{NS 5-brane}. In the 
absence of the orientifold, one would have 
${\rm SU}(N)\times {\rm SU}(N)$ with matter in the 
$(N, \bar N)\,\oplus\,(\bar N, N)$ representation. 
The orientifold ``identifies" the two ${\rm SU}(N)$ factors, 
projecting to the diagonal subgroup, giving a single ${\rm SU}(N)$ factor with 
one 
hypermultiplet in the antisymmetric representation for 
O$6^-$, or one hypermultiplet in the 
symmetric representation for O$6^+$ 
\cite{LandsteinerLopezLowe}.  It is important to note that the
orientifold  induces a ${\ZZ}_2$ involution in the curve. 
The NS 5-branes are at $x_7=x_8=x_9=0$, with {\it classically}
fixed values of $x_6$. The D4-branes have world-volume $x_0, x_1, x_2,
x_3, x_6$, 
with ends at fixed values of $x_6$, 
which gives a {\it macroscopic} world-volume on the D4-brane
as $d=4$. 
The M-theory picture gives rise to Riemann surface,  which is the SW curve
for this situation. 

\vskip1cm
\noindent{\bf 4. ~Hyperelliptic Perturbation Theory}
\renewcommand{\theequation}{4.\arabic{equation}}
\setcounter{equation}{0}

We have developed a systematic scheme for 
the instanton expansion for prepotentials 
associated to non-hyperelliptic curves \cite{oneanti}--\cite{santiago},  
which will be illustrated for the case of
${\rm SU}(N)$ gauge theory with one hypermultiplet in the antisymmetric
representation
\cite{oneanti}. The curve is given by (\ref{BBdos})
where $L^2\,=\,\Lambda^{N+2}$, with $\Lambda$ the 
quantum scale of the theory, and
\be
\epsilon =L^6 \,\,;\,\,\,\,\,\,\,\,\,\,\,\,
2A(x)&=&\left[ f(x)  + 3\,L^{2} \right]\,,\nn\\
f(x) = x^2 \prod^N_{i=1} (x-e_i)\,\,;\,\,\,\,\,\,\,\,\,\,\,\,
B(x)&=&L^{2} \left[ f(-x)  + 3\,L^{2} \right], \nn\\
{\rm involution:}\qquad y\to {L^4\over y},& & x\to -x.
\label{tfour}
\ee
It is fruitful to regard the last term 
$\epsilon= L^6$ in (\ref{BBdos}) as a perturbation. The 
intuition  is that this involves 
a much higher power of the quantum scale in 
(\ref{BBdos}) than the other terms, and 
geometrically it means separating the right-most 5-brane 
far from the remaining two \hbox{NS 5-branes}. 

To zeroth approximation we consider (\ref{BBdos}) with $\epsilon=0$, 
which is then a hyperelliptic curve, 
and can be analyzed by previously available methods
\cite{DHokerKricheverPhong1}--\cite{DHokerPhong}. This  approximation
gives ${\cal F}_{\rm 1-loop}$  correctly, but it is not adequate for
${\cal F}_{\rm 1-inst}$, so one needs to go beyond the hyperelliptic
approximation. We present  a systematic expansion in $\epsilon$, which is
not the same as an expansion in $L$, as the coefficient 
functions $A(x)$ and $B(x)$
depend on $L$. 
The perturbative expansion in $\epsilon$ for the solution   is 
\be
y_i =\bar y_i + \delta y_i = \bar y_i + 
\alpha_i \epsilon + \beta_i \epsilon^2 + \cdots,\qquad i=1,2,3,
\label{tnine}
\ee
which for our example gives to first order, 
\be
y_1(x) & = & -A-r-{L^6 (A-r)\over 2 B r}\,+\cdots\nn\\ 
y_2(x) & = & -A+r+{L^6 (A+r)\over 2 B r}\,+\cdots\nn\\  
y_3(x) & = & -{L^6 \over  B }\,+\cdots
\label{fone}
\ee
 with subscripts denoting the appropriate sheet, and with 
$r=\sqrt{A^2-B}$. It is straightforward to go to higher orders in
$\epsilon$.  

We note that sheet 3 is disconnected in any finite 
order of our perturbation expansion,
so we need only consider $y_1$ and $y_2$. 
We need the SW differential for 
these two sheets,   with the SW
differential for sheet 1
\be
\lambda_1 = x {dy_1 \over y_1},
\label{fseven}
\ee
and  $\lambda_2$ obtained from (\ref{fseven}) 
by $r \rightarrow {-r}$. Information on sheet 3 is obtained from the
involution symmetry in (\ref{tfour}). The expansion (\ref{tnine})  induces
a comparable expansion for
$\lambda$, whereby
\be
\lambda_1 &=&(\lambda_1)_I + (\lambda_1)_{II} + \cdots \,,
\label{feight}\\
\lambda_I
&=&  {x \left( {A^\prime\over A} - {B^\prime \over 2B} \right) \over 
\sqrt{1 - {B\over A^2}}}\,\, dx, \nn\\
\lambda_{II}
&=& -~ { L^6 \left(A - {B\over 2A}  \right)  \over
	 B^2 \sqrt{1 - {B\over A^2}} } \,\, dx, 
\label{fnine}
\ee
up to terms that do not contribute to period integrals. Notice that 
$\lambda_I$ is the SW differential obtained 
from the hyperelliptic approximation
($\epsilon=0$) to 
(\ref{BBdos}), and completely determines 
${\cal F}_{\rm 1-loop}$ and a part of the 1-instanton term, while
$\lambda_{II} \sim {\cal O}(L^2)$, so is of 
\hbox{1-instanton} order. Further
terms contribute 
only to \hbox{2-instanton} order and higher, so
(\ref{fnine}) is all that is needed  to \hbox{1-instanton} order.

In order to express the solutions to our problem 
with economical notation, we define 
 ``{\it residue functions}", $R_k(x)$, 
$S(x)$, $S_0(x)$, and $S_k(x)$, where
\be
{R_k(x)\over (x-e_k)} \,=\, {3\over f(x)},
\label{fifty}
\ee
and
\be
S(x)\,=\,{f(-x)\over f^2(x)}\,=\,{S_0(x)\over x^2}\,=\,
{S_k(x)\over (x-e_k)^2}.
\label{fione}
\ee

The functions $S(x)$ and $S_k(x)$  play a crucial role for understanding 
the general
features of the instanton expansion of SW problems. 

The \hbox{branch-cuts} are centered on the $e_i$ and connect
sheets $y_1$ and $y_2$ as shown in Fig.~2

\bigskip

\begin{picture}(500,50)(10,10)

\put(50,40){\line(1,0){60}}
\put(180,40){\line(1,0){60}}
\put(310,40){\line(1,0){60}}

\put(130,38){$\cdot$}
\put(135,38){$\cdot$}
\put(140,38){$\cdot$}
\put(145,38){$\cdot$}

\put(260,38){$\cdot$}
\put(265,38){$\cdot$}
\put(270,38){$\cdot$}
\put(275,38){$\cdot$}

\put(40,30){$x^-_1$}
\put(100,30){$x^+_1$}
\put(74,30){$e_1$}
\put(76,40){\circle*{2}}

\put(170,30){$x^-_k$}
\put(230,30){$x^+_k$}
\put(204,30){$e_k$}
\put(206,40){\circle*{2}}

\put(300,30){$x^-_N$}
\put(360,30){$x^+_N$}
\put(334,30){$e_N$}
\put(336,40){\circle*{2}}
\put(220,0){\makebox(0,0)[b]{\footnotesize\bf {Figure 2}}}
\end{picture}
\vskip.1in
The order parameters are computed in a 
canonical homology basis. For the order parameters $a_k$ we
have the basis $A_k$, as shown in Fig.~3,\vfil\eject

\begin{figure} [hbtp]
\begin{center}
\mbox{\psfig{file=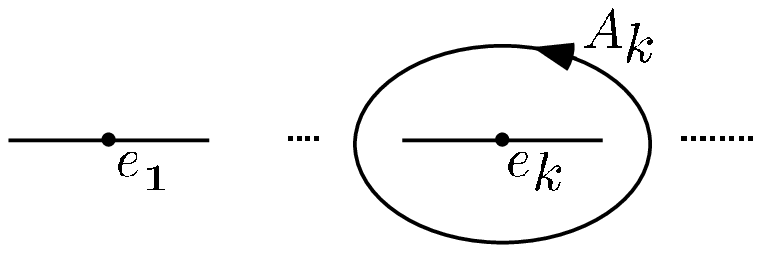,height=2cm}}
\end{center}
\end{figure}

\begin{center}
\vskip-.3in
{\footnotesize{\bf Figure 3}}
\end{center}

\noindent and the basis $B_k$ for the dual order parameters
$a_{D,k}$ as shown in Fig.~4,
\vskip.1in
\begin{figure} [hbtp]
\begin{center}
\mbox{\psfig{file=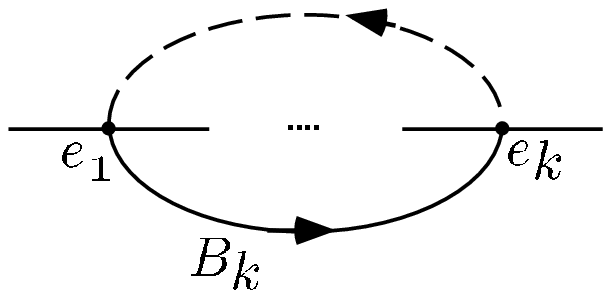,height=2cm}}
\end{center}
\end{figure}

\begin{center}
\vskip-.3in
{\footnotesize{\bf Figure 4}}
\end{center}

The cycle $B_k$ connects sheets $y_1$ and $y_2$, with
the solid line on sheet $y_1$ and the dashed line on
$y_2$, with $B_k$ passing through the \hbox{branch-cut} 
as shown. 
To compute (\ref{eight}), one only needs 
$(\lambda_1 - \lambda_2)$, 
so that we only need  terms odd 
under $r \rightarrow -r$. 
The order parameter is 
\be
2\pi{i} \, a_k & = & \oint_{A_k} \lambda  
	\simeq \oint_{A_k} \left(\lambda_I 
+ \lambda_{II} \,+\cdots \right) \nonumber \\
& = &
\oint_{A_k} dx \left[
\frac{x \left( \frac{A^\prime}{A} - \frac{B^\prime}{2B} \right)}
{\sqrt{ 1-\, \frac{B}{A^2}}} 
 - L^6 { \left(A - {B\over 2A}  \right)  \over
	 B^2 \sqrt{1 - {B\over A^2}} }
\right].
\label{fisix}
\ee
The second term does not contribute to ${\cal O}(L^2)$, 
as there are no poles at $x=e_k$. Thus to this order
\be
a_k = e_k + L^2 \left[ \frac{\pr S_k}{\pr x} (e_k)
- R_k (e_k) \right] + \cdots
\label{fieight}
\ee

The computation of the dual order parameter 
is considerably more complicated, 
with the result
\be
2 \pi i a_{D,k} &=& 2\pi i  (a_{D,k})_I\,+\, 2\pi i  (a_{D,k})_{II}\nn\\
 &=& 2 \pi i
{\pr \over  {\pr a_k}} \big[{\cal F}_{\rm classical} 
\,+\,{\cal F}_{\rm 1-loop}\big]\,
+\, L^2 \,{\pr \over {\pr a_k}} \big[- 2S_0(0)\,+
\,\sum_{k=1}^N S_k(a_k) \big],
\label{sifive}
\ee
so that  the one-instanton prepotential for ${\rm SU}(N)$ gauge theory
with one massless antisymmetric hypermultiplet
\cite{oneanti} turns out to be
\be
{\cal F}_{\rm 1-inst} 
 = {1\over 2\pi i}
 \left[ -2 \, S_0(0) +  \, \sum_k S_k (a_k) \right],
\label{siseven}
\ee
which is a prediction of \hbox{M-theory} that may be tested 
against microscopic calculations. 
This is presently possible for ${\rm SU}(N)$ with $N\leq 4$, since
${\rm SU}(2)\, +\,\Yasymm = {\rm SU}(2)$ (pure gauge theory); 
${\rm SU}(3)\, +\,\Yasymm = {\rm SU}(3) \,+$  $\fund$;
and  
${\rm SU}(4)\, +\,\Yasymm = {\rm SO}(6) \,+\,\fund $.
In each of these three cases,  (\ref{siseven})  
agrees with \hbox{1-instanton} calculations 
from ${\cal L}_{\rm micro}$ \cite{instanton}. 
For $N \geq  5$, (\ref{siseven})
should be regarded as predictions of \hbox{M-theory}, 
awaiting testing. The
fact that  (\ref{siseven})
agrees with microscopic calculations, when available, after 
a long derivation, with distinct methods, is already impressive. 

There are further applications of\,  hyperelliptic 
perturbation theory \cite{oneanti}--\cite{santiago}, where
the analysis is very similar to that 
sketched above. 

One may  add hypermultiplets 
in the fundamental representation, and  hypermultiplets with non-zero
masses.  For ${\rm SU}(N)$ gauge theory with an antisymmetric
representation and $N_f< N+2$, which is described by a cubic
SW curve \cite{LandsteinerLopezLowe}, one finds \cite{nonhyper}:

\be
2\pi i {\cal F}_{\rm 1-inst}=\sum_{k=1}^{N} S_k(a_k)-{2}S_m( -\tshalf m),
\label{seventy}
\ee
where 
\be
S_k(a_k)={(-1)^N\prod_{j=1}^{N_f}(a_k+M_j)\prod_{i=1}^{N}(a_k+a_i+m)\over 
(a_k+{1\over2}m)^2 \prod_{i\not=k}^N(a_k-a_i)^2},
\label{sevone}
\ee

\be
S_m(-\tshalf m)={(-1)^N\prod_{j=1}^{N_f}(M_j-{1\over2}m)\over 
\prod_{k=1}^N(a_k+{1\over2}m)},
\label{sevtwo}
\ee
where $M_j$ ($m$) is the mass of the hypermultiplet 
in the fundamental (resp. antisymmetric) 
representation. Eqs. (\ref{sevone}) and (\ref{sevtwo}) 
agree with scaling limits taking 
$M_j$ and/or $m\rightarrow\infty$. 
Eqs.~(\ref{seventy})--(\ref{sevtwo}) provide additional tests 
of \hbox{M-theory}, since ${\rm SU}(2)\, +\,\Yasymm\,+\, N_f
~\fund =  {\rm SU}(2)\,+N_f ~\fund$,  ${\rm
SU}(3)\, +\,\Yasymm\,+\, N_f ~\fund  = {\rm
SU}(3) \,+\, (N_f+1) ~\fund$, both of which agree with 
microscopic instanton calculations \cite{instanton}. 

An additional ${\rm SU}(N)$ theory with a cubic SW curve is 
${\rm SU}(N)\,+\,\Ysymm\,+\,N_f~ \fund$, 
with $N_f<N+2$. Here the result is
\cite{nonhyper}, \cite{product}:
\be
2\pi i {\cal F}_{\rm 1-inst}=\sum_{k=1}^{N} S_k(a_k),
\label{sevthree}
\ee
\be
S_k(a_k)={(-1)^N\,(a_k+{1\over2}m)^2\,
\prod_{j=1}^{N_f}(a_k+M_j)\,\prod_{i=1}^{N}(a_k+a_i+m)\over 
 \prod_{i\not=k}^N(a_k-a_i)^2}.
\label{sevfour}
\ee
This agrees with the microscopic calculation of Slater \cite{slater}.  

Whenever the predictions of the cubic SW curves 
obtained from \hbox{M-theory} have been tested, 
agreement has been found with those of microscopic calculations. 
However, there remain numerous further 
opportunities to subject \hbox{M-theory} predictions to testing.

\noindent{\bf 5. ~Universality}
\renewcommand{\theequation}{5.\arabic{equation}}
\setcounter{equation}{0}

If  one examines the cases discussed in the previous section, one
observes certain universal features: 

\noindent $(i)$ {The natural variables for this class of problems 
are the order parameters $\{a_k\}$ and 
not the 
gauge invariant moduli. 

\noindent $(ii)$ The \hbox{1-instanton} 
contribution to the prepotential can be written as 
\cite{onesym, nonhyper, DHokerKricheverPhong1}
\be
2\pi i {\cal F}_{\rm 1-inst}=\sum_{k=1}^{N} S_k(a_k),
\label{sevseven}
\ee
for ${\rm SU}(N)\,\,+\,\,N_f$ $\fund$ or ${\rm SU}(N)$ 
$+$ $\Ysymm$ $+$ $N_f$ $\fund$, and
\be
2\pi i {\cal F}_{\rm 1-inst}=\sum_{k=1}^{N} S_k(a_k)-2S_m(-\tshalf m),
\label{seveight}
\ee
for ${\rm SU}(N)$ $+$ $\Yasymm$ $+$ $N_f$ $\fund$
\cite{oneanti,nonhyper}.

Define  $S(x)$ which generalizes (\ref{fione}) as 
\be
S(x)\,=\,{S_k(x)\over (x-a_k)^2}
\,=\, {S_m(x)\over (x+\tshalf m)^2}.
\label{sevnine}
\ee
We tabulate the known $S(x)$ for ${\rm SU}(N)$ in the first 
three entries of Table $1$, where we
 include all generic 
cases of asymptotically 
free \hbox{${\cal N}=2$} ${\rm SU}(N)$ gauge theories. 
A careful examination of the first three rows of 
Table $1$ leads to the following empirical rules. 
$S(x)$ is given as the product of the 
following factors, each corresponding to 
a different \hbox{${\cal N}=2$} multiplet 
in a given representation of ${\rm SU}(N)$:

\noindent (1) Pure gauge multiplet factor 
\vglue-.3in
\be
{1\over \prod_{i=1}^{N}(x-a_i)^2}.
\label{eighty}
\ee

\noindent (2) Fundamental representation $\fund$. A factor 
\vskip-.15in
\be
(x+M_j)
\label{eione}
\ee
\vskip-.15in
\noindent
for each hypermultiplet of mass $M_j$ in the fundamental representation.

\noindent (3) Symmetric representation $\Ysymm$. A factor 
\be
(-1)^N\,(x+\tshalf m)^2\,\prod_{i=1}^{N}(x+a_i+m)
\label{eitwo}
\ee
for each hypermultiplet of mass $m$ in the symmetric representation.

\noindent (4) Antisymmetric representation $\Yasymm$. A factor 
\be
{(-1)^N\over(x+\tshalf m)^2}\,\prod_{i=1}^{N}(x+a_i+m)
\label{eithree}
\ee
for each hypermultiplet of mass $m$ in the antisymmetric representation.

{}From these empirical rules, we predict $S(x)$ for the last two
entries of Table~1. There are similar empirical rules for SO and Sp
\cite{elliptic}.

\begin{center} 
\begin{tabular}{||c|c||} 
\hline
\hline Hypermultiplet
Representations & $S(x)$\\ 
\hline\hline 
SU$(N)\,+\,N_f \, {\rm fund.}\, (M_j)$ &{} \\ [-.15in]
$(N_f \leq 2N)$ 
& ${\xf \prod_{j=1}^{N_f}(x+M_j)\over \prod_{i=1}^N (x-a_i)^2}$    \\
[-.15in] (ref. \cite{DHokerKricheverPhong1}) &{} \\ 
\hline
${\rm SU}(N)\,+\,1\,\,{\rm sym.}\,(m)  \,+\, N_f\,{\rm fund.}
 \, (M_j) $ &{}\\[-.15in]
$(N_f \,\leq N-2) $ 
& $ {\xf (-1)^N(x+\nha m)^2 \prod_{i=1}^N (x+a_i+\xxt m)
\prod_{j=1}^{N_f}(x+M_j) \over \prod_{i=1}^N (x-a_i)^2}$  \\[-.15in]
(ref. \cite{onesym,nonhyper}) &{} \\ 
\hline 
${\rm SU}(N)\,+\,1\,\,{\rm anti.} (m) \,+\, N_f\,{\rm fund.}\,(M_j)$ &{}
\\ [-.15in]
$(N_f\,\,\leq N+2)$ &
${\xf (-1)^N \prod_{i=1}^N (x+a_i+\xxt m) \prod_{j=1}^{N_f}(x+M_j) \over
(x+\nha m)^2 \prod_{i=1}^N (x-a_i)^2}$ \\ [-.15in]
(ref. \cite{oneanti,nonhyper}) & \\ 
\hline 
${\rm SU}(N)\, +\,2 \,\,{\rm anti.} \,(m_1,m_2) \,+\, N_f\,{\rm
fund.}\,(M_j)$  &{} \\[-.15in] 
$(N_f \,\,\leq 4)$ & ${\xf \prod_{i=1}^{N}(x+a_i+\xxt m_1) 
\prod_{i=1}^N(x
+a_i+\xxt m_2) \prod_{j=1}^{N_f}(x +M_j) \over (x +\nha m_1)^2 (x +\nha
m_2)^2
\prod_{i=1}^N (x -a_i)^2}$ \\[-.15in] (ref. \cite{twoanti}) 
&{} \\ 
\hline 
 &{} \\[-.2in]
SU$(N)$ + 1  anti. ($\xxt m_1$) + 1 sym. ($\xxt m_2$)  & 
${\xf  (x+\nha m_2)^2\,
\prod_{i=1}^N (x+a_i+\xxt m_1)  \prod_{i=1}^N (x+a_i+\xxt m_2) \over
(x+\nha m_1)^2 \prod_{i=1}^N (x-a_i)^2}$ \\[-.15in]
&{} \\
\hline \end{tabular} \end{center} \label{tableone}
\centerline{\footnotesize{\bf Table 1}:
The function $S(x)$
for SU$(N)$ gauge theory, with different matter content.}

Thus from these regularities we predict \cite{twoanti} for ${\rm SU}(N)$ +
2 $\Yasymm$ +$N_f\fund$ with $ N_f \le 4$:
\be
2\pi i {\cal F}_{\rm 1-inst}= 
\sum_{k=1}^N S_k(a_k) -2 S_{m_1}(-\tshalf m_1)-2S_{m_2}(-\tshalf m_2), 
\label{eifour}
\ee
where $S_k(a_k)$ and $S_m (-\tshalf m) $ 
are constructed from the  $4^{th}$ entry of Table $1$, 
even though 
no SW curve is available from \hbox{M-theory}! 

The predictions of Table 1 and (\ref{eifour}) can be tested as follows:

\noindent
1)  ${\rm SU}(2)\, +\,2 \,\Yasymm \,+ \,N_f \fund  = \,
     {\rm SU}(2)\,+\,N_f \fund, (N_f\leq 3)$.

\noindent
2)  ${\rm SU}(3)\, +\,2 \,\Yasymm + N_f \fund  = 
{\rm SU}(3) \,+\, (N_f+2)\fund, (N_f \leq 3)$. 

\noindent
3) Limit $m_1$ or $m_2 \, \rightarrow \infty$ 
reduces to ${\rm SU}(N) + \Yasymm+ N_f \fund$.  

In each of these three tests, our predicted ${\cal F}_{\rm 1-inst}$ 
finds agreement. However, it should be emphasized that there is no known
derivation of  the empirical rules of (\ref{eighty})-(\ref{eithree}). 
This is a problem that
deserves consideration from first principles. 

\noindent{\bf 6. ~Reverse Engineering a Curve}
\renewcommand{\theequation}{6.\arabic{equation}}
\setcounter{equation}{0}

Although there is no known SW curve for ${\rm SU}(N)$ gauge theory
with two antisymmetric
representations and $N_f\leq 3$ hypermultiplets, 
one can attempt to reverse
engineer a curve from the information in 
Table 1 and (\ref{eifour}). The strategy is 

\noindent
1) $ {\cal F}_{\rm classical}\,+\, {\cal F}_{\rm 1-loop}$ 
from perturbation theory. 

\noindent
2)  ${\cal F}_{\rm 1-inst}$ as predicted in Table 1 and  (\ref{eifour}).

\noindent
3) These two steps imply that 
$a_{D,k}= {{\pr {\cal F}} \over {\pr a_k}}$ 
is known to \hbox{1-instanton} accuracy. 

\noindent
4) Reproduce this expression from period 
integrals of a Riemann surface, to be
constructed from the above data.

\noindent
5) Ensure that the proposed Riemann surface 
is consistent with \hbox{M-theory}. 

Since  Witten has shown \cite{Witten}
that for ${\rm SU}(N)\times {\rm SU}(N)\times
\buildrel{m~{\rm factors}}\over{\cdots}\times \,
{\rm SU}(N)$ the corresponding curve is 
$$y^{m+1}+\cdots =0,$$
which results from $m+1$ parallel \hbox{NS 5-branes}, 
and $N$ \hbox{D4-branes} 
suspended between neighboring pairs of 
\hbox{NS 5-branes}. However, for $({\rm SU}(N))^m$, with $m\geq 3$, we 
have shown \cite{product} that to attain \hbox{1-instanton} accuracy, 
one {\sl only} needs a quartic approximation 
$$ y^4+\cdots=0,$$
to the full $y^{m+1}$ curve. Therefore, we only need a quartic curve 
to reproduce the 
prepotential to \hbox{1-instanton} accuracy. 
The most 
general quartic curve consistent with \hbox{M-theory} is of the form
\cite{Witten, twoanti}
\be
& & L^4\,j_1(x)\,P_2(x)\,t^2\,+\,L\,P_1(x)\,t\,+
\,P_0(x)\,+\,L\,j_0(x)\,P_{-1}(x)\,{1\over t}\nn\\
&  & +\,
L^4\,j_0{}^2(x)\,j_{-1}(x)\,\,P_{-2}(x)\,{1\over t^2}=0,
\label{eifive}
\ee
where $j_n(x)$ are associated to the $N_f$ flavors in the fundamental 
representation, and $P_n(x)$ are 
associated to the positions of \hbox{D4-branes}.

We  argue that  (\ref{eifive}) is incomplete if consistency with
\hbox{M-theory} is demanded, with the result of a curve of infinite order,
and therefore an infinite chain of \hbox{NS 5-branes} and orientifolds. 
To see the origin of this assertion, recall that the brane picture
for ${\rm SU}(N)$ gauge theory with an 
${\rm antisymmetric \,representation}$ 
of mass $m$ is shown in Fig.~1, and repeated in Fig.~5,
showing only the \hbox{NS 5-branes} and the O$6^-$
plane for clarity.
\vglue-.4in
\begin{picture}(400,150)(10,10)

\put(100,20){\line(0,1){75}}
\put(200,20){\line(0,1){75}}
\put(300,20){\line(0,1){75}}

\put(196.5,55){\footnotesize $\otimes$}
\put(204,60){\footnotesize O$6^-$}
\put(204,50){\footnotesize $(x+\tshalf m)$}
\put(220,0){\makebox(0,0)[b]{\footnotesize\bf {Figure 5}}}

\end{picture}
\vspace{.2in}

Therefore, for ${\rm SU}(N)$ gauge theory with two hypermultiplets in the 
antisymmetric representation 
of masses $m_1$ and $m_2$, 
we expect {\it at least} the brane structure in
Fig.~6.

\vglue-.4in
\begin{picture}(400,150)(10,10)

\put(80,20){\line(0,1){75}}
\put(170,20){\line(0,1){75}}
\put(250,20){\line(0,1){75}}
\put(330,20){\line(0,1){75}}

\put(166.5,55){\footnotesize$\otimes$}
\put(174,60){\footnotesize O$6^-$}
\put(174,50){\footnotesize$(x+\tshalf m_2)$}

\put(246.5,55){\footnotesize$\otimes$}
\put(254,60){\footnotesize O$6^-$}
\put(254,50){\footnotesize$(x+\tshalf m_1)$}
\put(220,0){\makebox(0,0)[b]{\footnotesize\bf {Figure 6}}}
\end{picture}

\vspace{.2in}

Again in  Fig.~6, only the \hbox{NS 5-branes} and O$6^-$
are shown, while the $N$ \hbox{D4-branes} connecting the \hbox{NS 5-branes} 
and the flavor \hbox{D6-branes}  are not shown for clarity. The
first observation is that to satisfy all possible mirrors,
one must have an infinite chain of \hbox{NS 5-branes} 
and O$6^-$ orientifolds, since one must satisfy the reflections
in {\it each}\, of the O$6^{-}$ orientifold planes
separately. A portion of this chain is shown in Fig.~7,
which differs from Fig.~6 in that the positions 
of \hbox{D4-branes} and D6-(flavor) branes are shown. 
One can check that all the necessary 
mirrors about any given O$6^-$
orientifold plane are satisfied.

Observe that if $m_2 \rightarrow \infty$, most of the \hbox{D4-branes},
\hbox{D6-branes} and O$6^-$ planes slide off to infinity, leaving 
us with the configuration 
of Fig.~5 for ${\rm SU}(N)$ and an antisymmetric representation of mass
$m_1$. 
Thus the infinite chain of \hbox{NS 5-branes} and O$6^-$ orientifolds,
a portion of which is shown in Fig.~7, yields
a curve of infinite order. 
The construction of such curves is described in the talk of S.G. Naculich
at this workshop. 

\begin{center}
\begin{picture}(810,295)(10,10)


\put(10,192){\line(1,0){5}}
\put(20,192){\line(1,0){5}}
\put(30,192){\line(1,0){5}}
\put(40,192){\line(1,0){5}}
\put(50,192){\line(1,0){5}}
\put(60,192){\line(1,0){5}}
\put(70,192){\line(1,0){5}}
\put(80,192){\line(1,0){5}}
\put(90,192){\line(1,0){5}}
\put(100,192){\line(1,0){5}}
\put(5,180){$(\!x\!-\!a_i\!+\!\xtw m_2\!-\!\xtw m_1\!)$}

\put(105,240){\line(1,0){2}}
\put(110,240){\line(1,0){5}}
\put(120,240){\line(1,0){5}}
\put(130,240){\line(1,0){5}}
\put(140,240){\line(1,0){5}}
\put(150,240){\line(1,0){5}}
\put(160,240){\line(1,0){5}}
\put(170,240){\line(1,0){5}}
\put(180,240){\line(1,0){5}}
\put(190,240){\line(1,0){5}}
\put(200,240){\line(1,0){5}}
\put(120,245){$(\!x\!+\!a_i\!+\!\xtw m_2\!)$}

\put(205,117){\line(1,0){2}}
\put(210,117){\line(1,0){5}}
\put(220,117){\line(1,0){5}}
\put(230,117){\line(1,0){5}}
\put(240,117){\line(1,0){5}}
\put(250,117){\line(1,0){5}}
\put(260,117){\line(1,0){5}}
\put(270,117){\line(1,0){5}}
\put(280,117){\line(1,0){5}}
\put(290,117){\line(1,0){5}}
\put(300,117){\line(1,0){5}}
\put(235,105){$(\!x-\!a_i\!)$}

\put(305,155){\line(1,0){2}}
\put(310,155){\line(1,0){5}}
\put(320,155){\line(1,0){5}}
\put(330,155){\line(1,0){5}}
\put(340,155){\line(1,0){5}}
\put(350,155){\line(1,0){5}}
\put(360,155){\line(1,0){5}}
\put(370,155){\line(1,0){5}}
\put(380,155){\line(1,0){5}}
\put(390,155){\line(1,0){5}}
\put(400,155){\line(1,0){5}}
\put(320,160){$(\!x\!+\!a_i\!+\!\xtw m_1\!)$}

\put(405,10){\line(1,0){2}}
\put(410,10){\line(1,0){5}}
\put(420,10){\line(1,0){5}}
\put(430,10){\line(1,0){5}}
\put(440,10){\line(1,0){5}}
\put(450,10){\line(1,0){5}}
\put(460,10){\line(1,0){5}}
\put(470,10){\line(1,0){5}}
\put(480,10){\line(1,0){5}}
\put(490,10){\line(1,0){5}}
\put(500,10){\line(1,0){5}}
\put(407,15){$(\!x\!-\!a_i\!+\!\xtw m_1\!-\!\xtw m_2\!)$}


\put(105,5){\line(0,1){255}}
\put(205,5){\line(0,1){255}}
\put(305,5){\line(0,1){255}}
\put(405,5){\line(0,1){255}}
\put(103,271){$\NSone$}
\put(203,271){$\NStwo$}
\put(303,271){$\NSthree$}
\put(403,271){$\NSfour$}


\put(101,215){$\otimes$} 
\put(201,175){$\otimes$}
\put(301,135){$\otimes$}
\put(401,85){$\otimes$}
\put(31,220){$(\!x\!+\!\xxt m_2\!-\!\nha m_1\!)$}
\put(160,170){$(\!x\!+\!\nha m_2\!)$}
\put(259,135){$(\!x\!+\!\nha m_1\!)$}
\put(406,75){$(\!x\!+\!\xxt m_1\!-\!\nha m_2\!)$}

\put(110,220){O$6^{-}$}
\put(210,180){O$6^{-}$}
\put(310,140){O$6^{-}$}
\put(410,90){O$6^{-}$}

\put(151,185){\framebox(5,5){$\cdot$}}
\put(251,165){\framebox(5,5){$\cdot$}}
\put(351,105){\framebox(5,5){$\cdot$}}
\put(120,194){$(\!x\!+\!\xxt m_2\!-\!M_j\!)$}
\put(235,172){$(\!x\!+\!M_j\!)$}
\put(320,96){$(\!x+\!\xxt m_1\!-\!M_j\!)$}

\put(30,30){\vector(1,0){20}}
\put(30,30){\vector(0,1){20}}
\put(25,50){$x$}
\put(50,25){$t$}

\end{picture}
\end{center}
\label{figureone}
{\footnotesize{\bf Figure 7}: \hfil\break
\noindent 1) {\it vertical lines} $|$: 
parallel, equally spaced \hbox{NS 5-branes}.

\noindent
2) {\it dashed lines} $- -$: 
$N$ parallel \hbox{D4-branes} connect pairs of adjacent \hbox{NS 5-branes}.

\noindent
3) $\otimes$: O$6^{-}$ orientifold planes.

\noindent
4) ${\framebox(5,5){$\cdot$}}\,$:  D6-(flavor) branes.

\noindent
Due to mirrors, the picture must extend infinitely to right and left.}
\vfil
\eject

\noindent{\bf 9. ~Concluding Remarks}

There are a number of open problems 
which should be addressed. An incomplete 
list is: 

\noindent
1) Compute ${\cal F}_{\rm 1-inst}$ from ${\cal L}_{\rm micro}$ for all 
the cases described in Table 1, 
so as to extend the test of \hbox{M-theory}.
In every case where a test can be made, agreement has been found. 

\noindent
2) Explain {\it group-theoretically} the entries
for $S(x)$ in Table 1, and the rules
 abstracted from these tables.

\noindent
3) Extend the predictions of non-hyperelliptic curves 
to regions of moduli space for which the hyperelliptic perturbation 
theory is not valid. 

\noindent
4) Enlarge the connections to integrable models.

As we have discussed, \hbox{${\cal N}=2$} SW theory presents 
many varied opportunities for testing \hbox{M-theory} predictions for
sypersymmetric gauge theories. These deserve to be explored further to
increase  our confidence in \hbox{M-theory}.

\centerline{\bf Acknowledgement}

We wish to thank the organizers of the workshop E. D'Hoker, D.H.
Phong, and S.T. Yau for the opportunity to present the results of our
group in such a stimulating atmosphere.

\end{document}